\documentstyle[12pt]{article}
\begin {document}

\title {THERMAL CONDUCTION IN SYSTEMS OUT OF HYDROSTATIC EQUILIBRIUM}
\author{L. Herrera\thanks{On leave from Departamento de F\'\i sica, 
Facultad de Ciencias, Universidad Central de Venezuela, Caracas, Venezuela 
and Centro de Astrof\'\i sica Te\'orica, M\'erida, Venezuela.}\ ,  
A. Di Prisco$^*$\ ,  
J. L. Hern\'andez-Pastora, 
J. Mart\'\i n\\
\'Area de F\'\i sica Te\'orica\\
Facultad de Ciencias\\
Universidad de Salamanca\\
37008, Salamanca, Espa\~na.\\
and
\and
J. Mart\'\i nez\\
Grupo de F\'\i sica Estad\'\i stica\\
Departamento de F\'\i sica\\
Universidad Aut\'onoma de Barcelona\\
08193 Bellaterra, Barcelona, Espa\~na.}
\date{}
\maketitle

\begin{abstract}
We analyse the effects of thermal conduction in a relativistic fluid, 
just after its departure from hydrostatic equilibrium, on a time 
scale of the order of thermal relaxation time. It is obtained 
that the resulting evolution will critically depend on a parameter 
defined in terms of thermodynamic variables, which is constrained by causality
 requirements

\end{abstract}

\section{Introduction.}

\noindent
Most part of the life of stars (at any stage of evolution), may be 
described on the basis of the quasi-static approximation (slowly 
evolving regime). This is so, because most relevant processes in star 
interiors take place on time scales that are usually, much larger 
than the hydrostatic time scale \cite{1},\cite{2}.

\noindent
However, during their evolution, self-gravitating objects may pass 
through phases of intense dynamical activity for which the quasi-static 
approximation is clearly not reliable (e.g., the quick collapse phase 
preceding neutron star formation). All these phases of star evolution 
(``slow'' and ``quick'') are generally accompanied by intense dissipative 
processes, usually described in the diffusion approximation. This 
assumption, in its turn, is justified by the fact that frequently, 
the mean free path of particles responsible for the propagation of 
energy in stellar interiors is very small as compared with the typical 
length of the star.

\noindent
In this work we shall study the influence of thermal conduction on the 
evolution of a self-gravitating system out of hydrostatic equilibrium 
(in the ``quick'' phase).

\noindent
However, instead of following its evolution long time 
after its departure from equilibrium, we 
shall evaluate the system immediately after such departure. Here 
``immediately'' means on a time scale of the order of thermal relaxation time,  
before the establishment of the steady state resistive flow.

\noindent
Doing so we shall avoid the introduction of numerical procedures  
which might lead to model dependent conclusions.

\noindent
On the other hand, however, we shall obtain only indications about 
the tendency of the object and not a complete description of its evolution.

\noindent
As we shall see, there appears a local parameter formed by a specific 
combination of thermal relaxation time, thermal conductivity, proper 
energy density and pressure, which critically affects the evolution 
of the object and which is constrained by causality requirements.

\noindent
The paper is organized as follows. 

\noindent
In the next section the 
field equations, the conventions and other useful formulae are introduced.
%\noindent
In section 3 we briefly present the equation for the heat conduction.
%\noindent
The central problem is analysed in section 4 and a discussion of 
results is given in the last section.

\section{Field Equations and Conventions.}

\noindent
We consider spherically symmetric distributions of collapsing 
fluid, which for sake of completeness we assume to be anisotropic, 
undergoing dissipation in the form of heat flow, bounded by a 
spherical surface $\Sigma$.

\noindent
The line element is given in Schwarzschild-like coordinates by
 
\begin{equation}
ds^2=e^{\nu} dt^2 - e^{\lambda} dr^2 - 
r^2 \left( d\theta^2 + sin^2\theta d\phi^2 \right)
\label{metric}
\end{equation}

\noindent
where $\nu(t,r)$ and $\lambda(t,r)$ are functions of their arguments. We 
number the coordinates: $x^0=t; \, x^1=r; \, x^2=\theta; \, x^3=\phi$.

\noindent
The metric (\ref{metric}) has to satisfy Einstein field equations
 
\begin{equation}
G^\nu_\mu=-8\pi T^\nu_\mu
\label{Efeq}
\end{equation}

\noindent 
which in our case read \cite{3}:

\begin{equation}
-8\pi T^0_0=-\frac{1}{r^2}+e^{-\lambda} 
\left(\frac{1}{r^2}-\frac{\lambda'}{r} \right)
\label{feq00}
\end{equation}

\begin{equation}
-8\pi T^1_1=-\frac{1}{r^2}+e^{-\lambda}
\left(\frac{1}{r^2}+\frac{\nu'}{r}\right)
\label{feq11}
\end{equation}

\begin{eqnarray}
-8\pi T^2_2  =  -  8\pi T^3_3 = & - &\frac{e^{-\nu}}{4}\left(2\ddot\lambda+
\dot\lambda(\dot\lambda-\dot\nu)\right) \nonumber \\
& + & \frac{e^{-\lambda}}{4}
\left(2\nu''+\nu'^2 - 
\lambda'\nu' + 2\frac{\nu' - \lambda'}{r}\right)
\label{feq2233}
\end{eqnarray}

\begin{equation}
-8\pi T_{01}=-\frac{\dot\lambda}{r}
\label{feq01}
\end{equation}

\noindent
where dots and primes stand for partial differentiation with respect
to t and r  
respectively.

\noindent
In order to give physical significance to the $T^{\mu}_{\nu}$ components 
we apply the Bondi approach \cite{3}.

\noindent
Thus, following Bondi, let us introduce purely locally Minkowski 
coordinates ($\tau, x, y, z$)

$$d\tau=e^{\nu/2}dt\,\qquad\,dx=e^{\lambda/2}dr\,\qquad\,
dy=rd\theta\,\qquad\, dz=rsin\theta d\phi$$

\noindent
Then, denoting the Minkowski components of the energy tensor by a bar, 
we have

$$\bar T^0_0=T^0_0\,\qquad\,
\bar T^1_1=T^1_1\,\qquad\,\bar T^2_2=T^2_2\,\qquad\,
\bar T^3_3=T^3_3\,\qquad\,\bar T_{01}=e^{-(\nu+\lambda)/2}T_{01}$$

\noindent
Next, we suppose that when viewed by an observer moving relative to these 
coordinates with velocity $\omega$ in the radial direction, the physical 
content  of space consists of an anisotropic fluid of energy density $\rho$, 
radial pressure $P_r$, tangential pressure $P_\bot$ and radial heat flux 
$\hat q$. Thus, when viewed by this moving observer the covariant tensor in 
Minkowski coordinates is

\[ \left(\begin{array}{cccc}
\rho    &  -\hat q  &   0     &   0    \\
-\hat q &  P_r      &   0     &   0    \\
0       &   0       & P_\bot  &   0    \\
0       &   0       &   0     &   P_\bot  
\end{array} \right) \]

\noindent
Then a Lorentz transformation readily shows that

\begin{equation}
T^0_0=\bar T^0_0= \frac{\rho + P_r \omega^2 }{1 - \omega^2} + 
\frac{2 Q \omega e^{\lambda/2}}{(1 - \omega^2)^{1/2}}
\label{T00}
\end{equation}

\begin{equation}
T^1_1=\bar T^1_1=-\frac{ P_r + \rho \omega^2}{1 - \omega^2} - 
\frac{2 Q \omega e^{\lambda/2}}{(1 - \omega^2)^{1/2}}
\label{T11}
\end{equation}

\begin{equation}
T^2_2=T^3_3=\bar T^2_2=\bar T^3_3=-P_\bot
\label{T2233}
\end{equation}

\begin{equation}
T_{01}=e^{(\nu + \lambda)/2} \bar T_{01}=
-\frac{(\rho + P_r) \omega e^{(\nu + \lambda)/2}}{1 - \omega^2} - 
\frac{Q e^{\nu/2} e^{\lambda}}{(1 - \omega^2)^{1/2}} (1 + \omega^2)
\label{T01}
\end{equation}

\noindent
with

\begin{equation}
Q \equiv \frac{\hat q e^{-\lambda/2}}{(1 - \omega^2)^{1/2}}
\label{defq}
\end{equation}

\noindent
Note that the velocity in the ($t,r,\theta,\phi$) system, $dr/dt$, 
is related to $\omega$ by

\begin{equation}
\omega=\frac{dr}{dt}\,e^{(\lambda-\nu)/2}
\label{omega}
\end{equation}

\noindent
At the outside of the fluid distribution, the spacetime is that of Vaidya, 
given by

\begin{equation}
ds^2= \left(1-\frac{2M(u)}{R}\right) du^2 + 2dudR - 
R^2 \left(d\theta^2 + sin^2\theta d\phi^2 \right)
\label{Vaidya}
\end{equation}

\noindent
where $u$ is a time-like coordinate such that $u=constant$ is
(asymptotically) a  
null cone open to the future and $R$ is a null coordinate
($g_{RR}=0$). It should  
be remarked, however, that strictly speaking, the radiation can be considered 
in radial free streaming only at radial infinity.

\noindent
The two coordinate systems ($t,r,\theta,\phi$) and ($u,R,\theta,\phi$) are 
related at the boundary surface and outside it by

\begin{equation}
u=t-r-2M\,ln \left(\frac{r}{2M}-1\right)
\label{u}
\end{equation}

\begin{equation}
R=r
\label{R}
\end{equation}

\noindent
In order to match smoothly the two metrics above on the boundary surface 
$r=r_\Sigma(t)$, we have to require the continuity of the first fundamental 
form across that surface. As result of this matching we obtain

\begin{equation}
\left[P_r\right]_\Sigma=\left[Q\,e^{\lambda/2}\left(1-\omega^2\right)^
{1/2}\right]_\Sigma = \left[\hat q\right]_\Sigma
\label{PQ}
\end{equation}

\noindent
expressing the discontinuity of the radial pressure in the presence 
of heat flow, which is a well known result \cite{4}.

\noindent
Next, it will be useful to calculate the radial components of the 
conservation law

\begin{equation}
T^\mu_{\nu;\mu}=0
\label{dTmn}
\end{equation}

\noindent
After tedious but simple calculations we get

\begin{equation}
\left(-8\pi T^1_1\right)'=\frac{16\pi}{r} \left(T^1_1-T^2_2\right) 
+ 4\pi \nu' \left(T^1_1-T^0_0\right) + 
\frac{e^{-\nu}}{r} \left(\ddot\lambda + \frac{\dot\lambda^2}{2}
- \frac{\dot\lambda \dot\nu}{2}\right)
\label{T1p}
\end{equation}

\noindent
which in the static case becomes

\begin{equation}
P'_r=-\frac{\nu'}{2}\left(\rho+P_r\right)+
\frac{2\left(P_\bot-P_r\right)}{r}
\label{Prp}
\end{equation}

\noindent
representing the generalization of the Tolman-Oppenheimer-Volkof equation 
for anisotropic fluids \cite{5}.

\section{Heat Conduction Equation.}

\noindent
As we mentioned in the introduction, in the study of star interiors 
it is usually assumed that the energy flux of radiation (and 
thermal conduction) is proportional to the gradient of temperature 
(Maxwell-Fourier law or Eckart-Landau in general relativity).

\noindent
However it is well known that the Maxwell-Fourier law for the radiation 
flux leads to a parabolic equation (diffusion equation) which predicts 
propagation of perturbation with infinite speed (see \cite{6}--\cite{8} and 
references therein). This simple fact is at the origin of the pathologies 
\cite{9} found in the approaches of Eckart \cite{10} and Landau \cite{11} 
for relativistic dissipative processes.

\noindent
To overcome such difficulties, different relativistic 
theories with non-vanishing relaxation times have been proposed 
in the past \cite{12}--\cite{15}. The important point is that all these 
theories provide a heat transport equation which is not of 
Maxwell-Fourier type but of Cattaneo type \cite{18}, leading thereby to a 
hyperbolic equation for the propagation of thermal perturbation.

\noindent
Accordingly we shall describe the heat transport by means of a 
relativistic Israel-Stewart equation \cite{8} , which reads 

\begin{equation}
\tau \frac{Dq^\alpha}{Ds} + q^\alpha = 
\kappa P^{\alpha \beta} \left(T_{,\beta} - T a_\beta\right) - 
\tau u^\alpha q_\beta a^\beta-
\frac{1}{2} \kappa T^2 
\left(\frac{\tau}{\kappa T^2} u^\beta\right)_{;\beta} q^\alpha
\label{Catrel}
\end{equation}

\noindent
where $\kappa$, $\tau$, $T$, $q^\beta$ and $a^\beta$ denote thermal
conductivity,  
thermal relaxation time, temperature, the heat flow vector and the
components of the four  
acceleration, respectively. Also, $P^{\alpha \beta}$ is the projector 
onto the hypersurface orthogonal to the four velocity $u^\alpha$.

\noindent
In our case this equation has only two independent components, 
which read, for $\alpha=0$

\[
\tau e^{(\lambda-\nu)/2}
\left(
Q \dot\omega + \dot Q \omega + Q \omega \dot\lambda
\right) +
\tau \left(
Q' \omega^2 + Q \omega \omega' + \frac{Q \omega^2 \lambda'}{2} 
\right)  
\]
\[
+ \frac{\tau Q  \omega^2}{r} 
+ Q \omega e^{\lambda/2} \left(1 - \omega^2\right)^{1/2}  =   
- \, \frac{\kappa \omega^2 \dot T e^{-\nu/2}}
{\left(1 - \omega^2\right)^{1/2}}
- \, \frac{\kappa \omega T' e^{-\lambda/2}}
{\left(1 - \omega^2\right)^{1/2}} 
\]
\[
- \, \frac{\nu'}{2} 
\frac{\kappa T \omega e^{-\lambda/2}}{\left(1 - \omega^2\right)^{1/2}}
- \frac12 Q \omega \left(e^{(\lambda-\nu)/2} \dot{\tau}+\omega \tau'\right)
\]
\[
-\frac12 \tau Q \omega \left[e^{(\lambda-\nu)/2}
\left(\frac{\omega \dot{\omega}}{1-\omega^2}+\frac{\dot{\lambda}}{2}\right)
+\left(\frac{\omega'}{1-\omega^2}+\frac{\nu'\omega}{2}\right)\right]
\]
\[
+\frac12 \tau Q \omega \left[\frac1\kappa
\left(e^{(\lambda-\nu)/2}\dot{\kappa}+\omega\kappa'\right)
+\frac2T\left(e^{(\lambda-\nu)/2}\dot{T}+\omega T'\right)
\right]
\]
\begin{eqnarray}
 + \left(\tau Q e^{(\lambda-\nu)/2} - \,  
\frac{\kappa T \omega e^{-\nu/2}}{\left(1 - \omega^2\right)^{1/2}}\right)
& \times &  
\left(\frac{\omega \dot\lambda}{2} + \frac{\dot\omega}{1 - \omega^2}\right) 
\nonumber \\ 
 + \left(\tau Q - \, 
\frac{\kappa T \omega e^{-\lambda/2}}{\left(1 - \omega^2\right)^{1/2}}\right)
& \times &
 \frac{\omega \omega'}{1 - \omega^2}
\label{com0}
\end{eqnarray}

\noindent
and for $\alpha=1$

\[
\tau e^{(\lambda-\nu)/2}
\left(
\dot Q + \frac{Q \dot\lambda}{2} + \frac{Q \omega^2 \dot\lambda}{2}
\right) + \tau \omega \left(
Q' + \frac{Q \lambda'}{2} \right) 
\]
\[
+\frac{\tau Q \omega}{r} + Q e^{\lambda/2} \left(1 - \omega^2\right)^{1/2} =
- \, \frac{\kappa \omega \dot T e^{-\nu/2}}
{\left(1 - \omega^2\right)^{1/2}}
- \, \frac{\kappa T' e^{-\lambda/2}}
{\left(1 - \omega^2\right)^{1/2}} 
\]
\[
-\, \frac{\nu'}{2}
\frac{\kappa T e^{-\lambda/2}}{\left(1 - \omega^2\right)^{1/2}}
-\, \frac{1}{2} Q \left(e^{(\lambda-\nu)/2}\dot{\tau}+\omega\tau'\right)
\]
\[
-\frac12 \tau Q \left[e^{(\lambda-\nu)/2}
\left(\frac{\omega \dot{\omega}}{1-\omega^2}+\frac{\dot{\lambda}}{2}\right)
+\left(\frac{\omega'}{1-\omega^2}+\frac{\nu'\omega}{2}\right)\right]
\]
\[
+\frac12 \tau Q \left[\frac1\kappa
\left(e^{(\lambda-\nu)/2}\dot{\kappa}+\omega\kappa'\right)
+\frac2T\left(e^{(\lambda-\nu)/2}\dot{T}+\omega T'\right)
\right]
\]
\begin{eqnarray}
 + \left(\tau Q \omega e^{(\lambda-\nu)/2} - \,  
\frac{\kappa T e^{-\nu/2}}{\left(1 - \omega^2\right)^{1/2}}\right)
& \times &  
\left(\frac{\omega \dot\lambda}{2} + \frac{\dot\omega}{1 - \omega^2}\right) 
\nonumber \\ 
 + \left(\tau Q \omega - \, 
\frac{\kappa T e^{-\lambda/2}}{\left(1 - \omega^2\right)^{1/2}}\right)
& \times &
 \frac{\omega \omega'}{1 - \omega^2}
\label{com1}
\end{eqnarray}

\noindent
where the expressions

\begin{equation}
u^\mu=\left(\frac{e^{-\nu/2}}{\left(1-\omega^2\right)^{1/2}},\,
\frac{\omega\, e^{-\lambda/2}}{\left(1-\omega^2\right)^{1/2}},\,0,\,0\right)
\label{umu}
\end{equation}

\begin{equation}
q^\mu=Q\,\left(\omega\,e^{(\lambda-\nu)/2},\,1,\,0,\,0\right)
\label{qmu}
\end{equation}

\noindent
have been used.

\noindent
We are now ready to get into the central problem of this work.

\section{Thermal Conduction and Departure from Hydrostatic Equilibrium.}

\noindent
Let us now consider a spherically symmetric fluid distribution which 
initially may be in either hydrostatic and thermal equilibrium (i.e. 
$\omega = Q = 0$), or slowly evolving and dissipating energy through 
a radial heat flow vector.

\noindent
Before proceeding further with the treatment of our problem, let us 
clearly specify the meaning of ``slowly evolving''. That means that 
our sphere changes on a time scale which is very large as compared to 
the typical time in which it reacts on a slight perturbation of 
hydrostatic equilibrium. This typical time is called hydrostatic 
time scale. Thus a slowly evolving system is always in hydrostatic 
equilibrium (very close to), and its evolution may be regarded as 
a sequence of static models linked by (\ref{feq01}).

\noindent
As we mentioned before, this assumption is very sensible, since 
the hydrostatic time scale is usually very small.

\noindent
Thus, it is of the order of $27$ minutes for the sun, $4.5$ seconds 
for a white dwarf and $10^{-4}$ seconds for a neutron star of one 
solar mass and $10$ Km radius \cite{2}.

\noindent
In terms of $\omega$ and metric functions, slow evolution means 
that the radial velocity $\omega$ measured by the Minkowski observer, 
as well as time derivatives are so small that their products and 
second order time derivatives may be neglected (an invariant 
characterization of slow evolution may be found in \cite{16}).

\noindent
Thus \cite{17}

\begin{equation}
\ddot\nu\approx\ddot\lambda\approx\dot\lambda \dot\nu\approx
\dot\lambda^2\approx\dot\nu^2\approx
\omega^2\approx\dot\omega=0
\label{neg}
\end{equation}

\noindent
As it follows from (\ref{feq01}) and (\ref{T01}), $Q$ is of the 
order $O(\omega)$. 
Thus in the slowly evolving regime, relaxation terms may be neglected 
and (\ref{Catrel}) becomes the usual Landau-Eckart transport equation 
\cite{17}.

\noindent
Then, using (\ref{neg}) and (\ref{T1p}) we obtain (\ref{Prp}), 
which as mentioned before is the equation of hydrostatic equilibrium 
for an anisotropic fluid. This is in agreement with what was mentioned 
above, in the sense that a slowly evolving system is in hydrostatic 
equilibrium.

\noindent
Let us now return to our problem. Before perturbation, the two 
possible initial states of our system are characterized by:

\begin{enumerate}
\item Static
\begin{equation}
\dot \omega = \dot Q = \omega = Q = 0 
\label{eqdt}
\end{equation}
\item Slowly evolving
\begin{equation}
\dot \omega = \dot Q = 0
\label{evlen}
\end{equation}
\begin{equation}
Q \approx O(\omega) \not = 0 \; \qquad (small)
\label{Qorom}
\end{equation}
\end{enumerate}

\noindent
where the meaning of ``small'' is given by (\ref{neg}).

\noindent
Let us now assume that our system is submitted to perturbations 
which force it to depart from hydrostatic equilibrium but keeping the 
spherical symmetry.

\noindent
We shall study the perturbed system on a time scale which is 
small as compared to the thermal adjustment time.

\noindent
Then, immediately after perturbation (``immediately'' understood 
in the sense above), we have for the first initial condition 
(static) 

\begin{equation}
\omega = Q = 0
\label{omyQ0}
\end{equation}

\begin{equation}
\dot\omega \approx \dot Q \not = 0 \; \qquad (small)
\label{chiq}
\end{equation}

\noindent
whereas for the second initial condition (slowly evolving)

\begin{equation}
Q \approx O(\omega) \not = 0 \; \qquad (small)
\label{Qseg}
\end{equation}

\begin{equation}
\dot Q \approx \dot\omega \not = 0 \; \qquad (small)
\label{pomQ2}
\end{equation}

\noindent
As we shall see below, both initial conditions lead to the same final 
equations.

\noindent
Let us now write explicitly eq.(\ref{T1p}). With the help of 
(\ref{T00})--(\ref{T01}), we find after long but trivial calculations 

\begin{eqnarray}
& & \frac{P_r'}{1-\omega^2} + 
\frac{\rho' \omega^2}{1-\omega^2} + 
\frac{2 \omega \omega' \rho}{1-\omega^2} + 
\frac{2 \omega \omega' P_r}{\left(1-\omega^2\right)^2}   
\nonumber \\ 
& + &
\frac{2 \omega^3 \omega' \rho}{\left(1-\omega^2\right)^2}  +
\frac{2 Q' \omega e^{\lambda/2}}{\left(1-\omega^2\right)^{1/2}} +
\frac{2 Q \omega' e^{\lambda/2}}{\left(1-\omega^2\right)^{1/2}} + 
\frac{2 Q \omega^2 \omega' e^{\lambda/2}}{\left(1-\omega^2\right)^{3/2}}  
\nonumber \\
& + &
\frac{2}{r} \, [ \, 
\frac{4 \pi r^3}{r-2m} \, 
\left(\rho + P_r \omega^2\right) \, 
\frac{Q \omega e^{\lambda/2}}{\left(1-\omega^2\right)^{3/2}} 
+ \frac{12 \pi r^3}{r-2m} \, 
\left(\frac{Q \omega e^{\lambda/2}}{\left(1-\omega^2\right)^{1/2}}\right)^2   
\nonumber \\
& + & 
\left(\rho + P_r\right) \, \frac{\omega^2}{1-\omega^2} 
+ \left(P_r - P_\bot\right)  + 
\frac{2 Q \omega e^{\lambda/2}}{\left(1-\omega^2\right)^{1/2}} +  
\frac{\left(\rho+P_r\right)}{2} \, 
\frac{1+\omega^2}{1-\omega^2} \, \frac{m}{r-2m} 
\nonumber \\ 
& + &
\frac{Q \omega e^{\lambda/2}}{\left(1-\omega^2\right)^{1/2}} \, 
\frac{m}{r-2m} + 
\frac{2 \pi r^3}{r-2m} \, 
\left(P_r+\rho\omega^2\right) \left(\rho+P_r\right) \, 
\frac{1+\omega^2}{\left(1-\omega^2\right)^2} 
\nonumber \\ 
& + &
\frac{8 \pi r^3}{r-2m} \, 
\left(P_r+\rho\omega^2\right) \, 
\frac{ Q \omega e^{\lambda/2}}{\left(1-\omega^2\right)^{3/2}}  + 
\frac{4 \pi r^3}{r-2m} \, 
Q \omega e^{\lambda/2} \left(\rho+P_r\right) \, 
\frac{1+\omega^2}{\left(1-\omega^2\right)^{3/2}} \, ] 
\nonumber \\
& = &
\frac{e^{-\nu}}{8 \pi r} 
\left(\ddot\lambda + 
\frac{\dot\lambda^2}{2} - 
\frac{\dot\lambda \dot\nu}{2}\right)
\label{horror}
\end{eqnarray}

\noindent
which, when evaluated immediately after perturbation, reduces to

\begin{equation}
P'_r + \frac{\left(\rho + P_r\right) m}{r^2 \left(1 - 2m/r\right)} 
+ \frac{4 \pi r}{\left(1 - 2m/r\right)} \left(P_r \rho + P_r^2\right) 
+ \frac{2 \left(P_r - P_\bot\right)}{r} 
= \frac{e^{- \nu}}{8 \pi r} \ddot \lambda
\label{menho}
\end{equation}

\noindent
for both initial states.

\noindent
On the other hand, an expression for $\ddot\lambda$ may be obtained by 
taking the time derivative of (\ref{feq01})

\begin{eqnarray}
\ddot\lambda & = & -  8 \pi r e^{(\nu + \lambda)/2} 
[\,
\left(\rho + P_r\right) \frac{\omega}{1-\omega^2} 
\frac{\dot\nu}{2} + 
Q e^{\lambda/2} \frac{1+\omega^2}{\left(1-\omega^2\right)^{1/2}} 
\frac{\dot\nu}{2} \nonumber \\
& + & 
\frac{\left(\rho+P_r\right) \omega}{1-\omega^2} 
\frac{\dot\lambda}{2} +  
Q e^{\lambda/2} \frac{1+\omega^2}{\left(1-\omega^2\right)^{1/2}} 
\dot\lambda + 
\left(\dot\rho + \dot P_r\right) 
\frac{\omega}{1-\omega^2} \nonumber \\
& + &
\left(\rho+P_r\right) \dot\omega 
\frac{1+\omega^2}{\left(1-\omega^2\right)^{2}}  + 
\dot Q e^{\lambda/2} \frac{1+\omega^2}{\left(1-\omega^2\right)^{1/2}} 
\nonumber \\ 
& + &
Q e^{\lambda/2} \frac{\omega \dot\omega \left(3-\omega^2\right)}
{\left(1-\omega^2\right)^{3/2}}
\,]
\label{pplex}
\end{eqnarray}

\noindent
which, in its turn, when evaluated after perturbation, reads

\begin{equation}
\ddot\lambda = - 8 \pi r e^{(\nu+\lambda)/2} 
\left[\left(\rho+P_r\right) \dot\omega + 
\dot Q e^{\lambda/2}\right]
\label{ddl12}
\end{equation}

\noindent
replacing $\ddot \lambda$ by (\ref{ddl12}) en (\ref{menho}), 
we obtain

\begin{equation}
- e^{(\nu-\lambda)/2} R = \left(\rho+P_r\right) \dot\omega + 
\dot Q e^{\lambda/2}
\label{pfR}
\end{equation}

\noindent
where $R$ denotes the left-hand side of the TOV equation, i.e.

\begin{eqnarray}
R & \equiv &  \frac{dP_r}{dr} + \frac{4\pi r P_r^2}{1-2m/r} + 
\frac{P_r m}{r^2 \left(1-2m/r\right)} + 
\frac{4\pi r \rho P_r}{1-2m/r} + \nonumber \\ 
 &  & + \frac{\rho m}{r^2 \left(1-2m/r\right)} - 
\frac{2\left(P_\bot - P_r\right)}{r} \nonumber \\
& = & P'_r + \frac{\nu'}{2} \left(\rho + P_r\right) -
\frac{2}{r} \left(P_\bot - P_r\right) 
\label{Rfr}
\end{eqnarray}

\noindent
The physical meaning of $R$ is clearly inferred from (\ref{Rfr}). 
It represents the total force (gravitational + pressure gradient + 
anisotropic term) acting on a given fluid element. Obviously, 
$R>0/R<0$ means that the total force is directed $inward/outward$ of 
the sphere.

\noindent
Let us now turn back to thermal conduction equation (\ref{Catrel}). 
Evaluating its $t$-component (given by Eq.(\ref{com0}))
immediately after perturbation, we obtain for the first initial 
configuration (static), an identity. Whereas the second case 
(slowly evolving) leads to

\begin{equation}
\omega \left(T' + T \frac{\nu'}{2}\right) = 0
\label{cs2}
\end{equation}

\noindent
which is to be expected, since before perturbation, in the 
slowly evolving regime, we have according to Eckart-Landau 
(valid in this regime)

\begin{equation}
Q = - \kappa e^{-\lambda} \left(T' + \frac{T \nu'}{2}\right)
\label{EL}
\end{equation}

\noindent
Therefore, the quantity in bracket is of order $Q$. Then 
immediately after perturbation this quantity is still of 
order $O(\omega)$, which implies (\ref{cs2}).

\noindent
The corresponding evolution of the $r$-component of the equation 
(\ref{Catrel}) yields, for the initially static configuration

\begin{equation}
\tau \dot Q e^{\lambda/2} = - \kappa T \dot\omega 
\label{Cat1}
\end{equation}

\noindent
where the fact has been used that after perturbation

\begin{equation}
Q = 0 \quad \Longrightarrow \quad T' = - \, \frac{T \nu'}{2}
\label{impl}
\end{equation}

\noindent
For the second case, the $r$-component of heat transport equation 
yields also (\ref{Cat1}), since after perturbation the value of $Q$ 
is still given by (\ref{EL}), up to $O(\omega)$ terms.

\noindent
Finally, combining (\ref{pfR}) and (\ref{Cat1})

\begin{equation}
\dot\omega = - \frac{e^{(\nu-\lambda)/2} R}{\left(\rho+P_r\right)} 
\times 
\frac{1}{\left(1 - \frac{\kappa T}{\tau \left(\rho+P_r\right)}\right)}
\label{exmin}
\end{equation}

\noindent
or, defining the parameter $\alpha$ by 

\begin{equation}
\alpha \equiv \frac{\kappa T}{\tau \left(\rho + P_r\right)}
\label{alfa}
\end{equation}

\begin{equation}
- e^{(\nu-\lambda)/2} R = 
\left(\rho + P_r\right) \dot \omega \left(1 - \alpha\right)
\label{Ralfa}
\end{equation}

\noindent
Let us first consider the $\alpha=0$ case. Then, last expression 
has the obvious ``Newtonian'' form

\centerline{Force $=$ mass $\times$ acceleration}

\noindent
since, as it is well known, $\left(\rho + P_r\right)$ represents 
the inertial mass density and by ``acceleration'' we mean the time derivative 
of $\omega$ and not $(a_\mu a^\mu)^{1/2}$. In this case ($\alpha=0$), an 
$outward/inward$ acceleration ($\dot \omega>0/\dot \omega<0$) is 
associated with an $outwardly/inwardly$ ($R<0/R>0$) directed total 
force (as one expects!). 

\noindent
However, in the general case ($\alpha \not = 0$) the situation 
becomes quite different. Indeed, the inertial mass term is now 
multiplied by ($1-\alpha$), so that if $\alpha=1$, we obtain that 
$\dot \omega \not = 0$ even though $R=0$. Still worse, if 
$\alpha>1$, then an $outward/inward$ acceleration is associated with an 
$inwardly/outwardly$ directed total force!.However as we shall see in
 next section,causality requirements constrain $\alpha$ to be less than 1.

\noindent
The last term in (\ref{Catrel}) is frequently omitted (the so-called 
``truncated'' theory) \cite{19}. In the context of this work both 
components of this term vanish and therefore all results found above 
are independent of the adopted theory (Israel-Stewart or truncated).

\section{Discussion.}

\noindent
Restrictions based on stability and causality were derived in \cite{9} for
Israel-Stewart thermodynamics.According to equation (134) of \cite{9}, 
it follows that we must have $\alpha<1$ in order to guarantee that thermal
pulses are subluminal

\noindent
In fact there is a similar parameter in the case of bulk viscous
perturbations  
\cite{8}, which due to causality an stability limits should be
smaller than one.  

\noindent
Before concluding we would like to make the following remarks:
\begin{enumerate}
\item Observe the formal similarity between the critical point and 
the equation of state for an inflationary scenario ($\rho = - P_r$).
\item It should be clear that in the context of the perturbation scheme 
used here, we get information only about the tendency of the system. To 
find out the real influence of the critical point on the evolution of 
the object, the full integration of the equations is required. Calculations 
involving such integrations have been performed in the past 
\cite{26}--\cite{28}, however in neither one of the examples examined there, 
the system reaches the critical point. Furthermore, our configurations are 
initially in global thermal equilibrium, which is a highly idealized 
situation. In this sense, it should be stressed that our aim here is not to 
model a real star but to study some specific aspects of relativistic
diffusion.
\item It should be clear that the analysis presented here depends strictly 
on the validity of the diffusion approximation, which in turn depends on 
the assumption of local thermodynamical equilibrium (LTE). Therefore, only 
small deviations from LTE can be considered in the context of this work.
\item For the sake of completeness we have considered an anisotropic fluid 
(instead of an isotropic one), leaving the origin of such anisotropy
completely  
unspecified. As it is apparent, anisotropy does not affect the most important 
result obtained here (Eq.(\ref{Ralfa})). However, should anisotropy be related 
to viscosity, then for consistency the anisotropic pressure tensor should be 
subjected to the Israel-Stewart causal evolution equation for shear viscosity
\end{enumerate} 

\section*{Acknowledgments.}

\noindent
This work has been partially supported by the Spanish Ministry of
Education under grant No. PB94-0718.

\end{document}